\documentclass[preprint,12pt]{elsarticle}
\usepackage{amsmath,amssymb,amsfonts}
\usepackage{algorithmic}
\usepackage{algorithm}
\usepackage{array}
\usepackage{subfig}
\usepackage{textcomp}
\usepackage{url}
\usepackage{verbatim}
\usepackage{graphicx}
\usepackage{xcolor}
\usepackage{float}
\usepackage{bm}

\begin{document}
	
	\begin{frontmatter}
		\title{Digital Self-Interference Cancellation for In-Band Full-Duplex Systems Based on Transformer Neural Operators}
		\author[scu]{Zhijie Su}
		\ead{zhijiesu164@163.com}
		\author[scu]{Shiquan Zhang\corref{cor1}}
		\ead{shiquanzhang@scu.edu.cn}
		\author[scu]{Qiaolin He\corref{cor1}}
		\ead{qlhejenny@scu.edu.cn}
		\cortext[cor1]{Corresponding authors.}
		\fntext[funding]{This research is partially supported by the National Key R\&D Program of China (No. 2022YFE03040002) and the National Natural Science Foundation of China (No. 12371434).}
		\address[scu]{Department of Mathematics, Sichuan University, Chengdu 610065, China}
		\begin{abstract}
			Digital-domain self-interference cancellation is essential in in-band full-duplex radios after passive suppression and analog cancellation. It is particularly challenging when the desired signal is strong relative to the residual self-interference. In this regime, conventional iterative digital cancellers update their coefficients from desired-signal-contaminated received samples and may suffer substantial performance degradation. This paper reformulates digital-domain self-interference cancellation as an operator-learning problem. The proposed framework learns the temporal evolution of the residual self-interference channel from historical observations and uses the learned operator together with the known local transmit signal to reconstruct the current residual self-interference. A Transformer neural operator is used to realize this framework. The resulting method is trained in a desired-signal-free setting and then applied in the presence of a desired signal. A continuity-based analysis characterizes the sensitivity of the learned operator to the resulting input perturbation. Numerical experiments in static and time-varying channels, including time-varying SIR conditions, demonstrate stable cancellation performance over a wide range of interference levels. At 30 dB SIR, the proposed method improves the cancellation performance by 5--15 dB over the considered conventional methods.
		\end{abstract}
		
	\end{frontmatter}

	\section{Introduction}
	In modern wireless communication, the demand for higher data rates, larger network capacity, and lower latency continues to grow. Conventional half-duplex systems separate transmission and reception in time or frequency and are increasingly unable to meet these requirements. Full-duplex communication enables simultaneous transmission and reception over the same frequency band \cite{Full-Duplex1,Full-Duplex2,Full-Duplex3,Full-Duplex4} and can potentially double spectral efficiency.
	
	A major obstacle to full-duplex communication is self-interference \cite{self-interference1,self-interference2,self-interference3,self-interference4}. A portion of the local transmit signal leaks or is reflected into the local receiver together with the desired signal sent by the remote node. This SI can severely impair desired-signal decoding. Bharadia \emph{et al.} \cite{multi-sic} demonstrated a practical three-stage architecture that combines antenna isolation, analog cancellation, and digital cancellation. This paper focuses specifically on digital-domain cancellation of the residual SI that remains after passive suppression and analog cancellation.
	
	Digital SI cancellation faces two primary challenges. First, SI paths caused by leakage and reflections can vary over time and must be tracked accurately. Second, the desired signal acts as interference when the residual SI channel is estimated from the received signal. When previous cancellation stages are effective, the desired signal may be significantly stronger than the residual SI, which makes subsequent digital cancellation difficult \cite{digital-sic,LMS3}.
	
	LMS is one of the most widely used digital SI cancellation methods \cite{LMS1,LMS2,LMS3}. Related approaches include NLMS \cite{NLMS1,NLMS2}, Sign-LMS \cite{Sign-LMS}, variable-step-size LMS \cite{VSS-LMS1,VSS-LMS2,VSS-LMS3}, conjugate-gradient methods \cite{CG}, and RLS \cite{RLS}. Neural-network-based adaptive filtering has also been investigated \cite{NN-base1,NN-base2,NN-base3}, providing data-driven alternatives to conventional adaptive filters. Although these methods can be effective in some settings, their performance may deteriorate when the desired signal is strong relative to the SI component being estimated.
	
	Nonlinear SI models provide one way to improve cancellation performance. Komatsu \emph{et al.} \cite{theory-full-duplex} showed that nonlinear cancellation models can outperform conventional linear models in full-duplex radios. Memory-polynomial nonlinear cancellers have subsequently been proposed to model transmitter-chain nonlinearities \cite{NONLMS1,NONLMS2}.
	
	Existing methods mainly improve step-size selection, search directions, or nonlinear basis functions. They nevertheless estimate cancellation parameters by iterating on received observations. Consequently, their coefficient updates remain sensitive to desired-signal contamination, especially in high-SIR settings.
	
	To address this limitation, we consider a data-driven framework that directly characterizes the relation among the local transmit signal, SI-channel evolution, and the resulting SI. Since SI is determined by the local transmit signal and the SI channel, rather than by the desired signal, a learned operator that predicts the current channel response from historical channel behavior can reduce reliance on contaminated instantaneous updates.
	
	Operator learning has recently attracted attention for learning mappings between function spaces. Representative architectures include DeepONet \cite{DeepOnet,Shift-DeepOnet}, Transformer-based neural operators \cite{transformer,transformer2}, physics-informed and convolutional neural operators \cite{PINO,CNO}, and Fourier-based neural operators \cite{FNO,FAN}.
	
	A recent closely related study recasts digital SI calibration as an in-context regression problem \cite{Balaji2026}. A Transformer is pre-trained over a distribution of transmitter, channel, and front-end conditions. Given a small set of pilot-based calibration examples from the current operating state, it directly predicts the residual SI through a single forward pass, without re-estimating filter coefficients or performing online gradient-based fine-tuning. The context examples and the query are assumed to belong to the same stationary radio state; thus, the method is designed for few-shot calibration within a quasi-static operating interval.

	In contrast, this paper reformulates full-duplex SI cancellation as an operator-learning problem for temporal SI-channel evolution, where the Transformer neural operator is one realization of the operator-learning framework rather than the contribution by itself. The proposed formulation targets desired-signal contamination at high SIR and rapidly varying SI channels. It exploits ordered historical received observations and periodic desired-signal-free calibration data to fine-tune the learned operator, thereby continuously tracking SI channels whose variations are independent of those of the desired-signal strength. The contributions are summarized as follows.
	
	\begin{enumerate}
		\item The SI cancellation problem is reformulated as an operator-learning problem that maps historical SI-channel behavior and the known transmit signal to the current SI component.
		\item A Transformer-neural-operator-based SI cancellation method is developed within this framework. An input-perturbation analysis characterizes sensitivity to desired-signal contamination under a small desired-to-local-transmit-signal condition and attributes this property to the proposed clean-training/contaminated-inference formulation rather than to a particular attention mechanism alone.
		\item Numerical experiments cover different SIR levels and static and time-varying SI channels. In the static-channel SIR study, the largest pairwise difference in cancellation performance among all tested SIR values is no greater than 0.05 dB; in the dynamic-channel experiments, the proposed method achieves a 5--15 dB gain over the considered conventional methods at 30 dB SIR.
	\end{enumerate}
	
	The remainder of this paper is organized as follows. Section~\ref{sec:model} describes the SI cancellation model. Section~\ref{sec:method} presents the digital self-interference cancellation for in-band
	full-duplex systems based on transformer neural
	operators and its analysis. Section~\ref{sec:Numerical Experiments} reports the numerical experiments, and Section~\ref{sec:conclusion} concludes the paper.

	\section{Self-Interference Cancellation Model}
	\label{sec:model}
	\subsection{Notation and statistical assumptions}
	Throughout this paper, $x(n)$ and $s(n)$ denote realizations of complex-valued discrete-time stochastic processes: $x(n)$ is the known local transmit signal and $s(n)$ is the desired signal transmitted by the remote node. The SI channel $h_n$ is an unknown channel mapping that may vary with $n$; it is fixed in the static experiments and time varying in the dynamic experiments. Conditional on $x(n)$ and $h_n$, the SI signal $l(n)$ is deterministic. We assume that $x(n)$ and $s(n)$ have finite second moments and are uncorrelated,
	\begin{equation}
		\mathbb{E}\{X^*(n)s(n)\}=\bm{0},
		\label{eq:statistical-assumptions}
	\end{equation}
	where $X(n)=[x(n-N+1),x(n-N+2),\ldots,x(n)]$ is the local transmit-signal vector. The expectation in \eqref{eq:statistical-assumptions} is an ensemble expectation; it does not imply $X^*(n)s(n)=\bm{0}$ for every observed time instant. Receiver noise is omitted in the following notation for clarity, as in the original signal model; it can be included additively in $r(n)$ without changing the argument.
	
	\subsection{Self-interference model and instantaneous optimization problems}
	The received signal is
	\begin{equation}
		r(n)=s(n)+l(n),
		\label{eq:r(n)}
	\end{equation}
	where the SI component is generated by the current and previous transmit samples,
	\begin{equation}
		l(n)=h_n\bigl(X(n)\bigr), \qquad h_n:\mathbb{C}^{N}\rightarrow\mathbb{C}.
		\label{eq:l(n)2}
	\end{equation}
	A linear canceller uses $d(n)=X(n)W$, where $W\in\mathbb{C}^{N}$ is the tap-coefficient vector. The ideal cancellation error and the observable residual are, respectively,
	\begin{equation}
		err(n;W)=l(n)-X(n)W,
		\label{eq:err(n)}
	\end{equation}
	and
	\begin{equation}
		e(n;W)=r(n)-X(n)W=err(n;W)+s(n).
		\label{eq:e(n)}
	\end{equation}
	
	The problem considered in this paper is explicitly a sequence of instantaneous optimization problems indexed by $n$. At time $n$, the ideal but unobservable problem is
	\begin{equation}
		\mathcal{P}^{\mathrm{SI}}_n:\qquad
		\min_{W}\ |err(n;W)|^2,
		\label{eq:ideal-problem}
	\end{equation}
	whereas the receiver can only form
	\begin{equation}
		\mathcal{P}^{\mathrm{rx}}_n:\qquad
		\min_{W}\ |e(n;W)|^2.
		\label{eq:observable-problem}
	\end{equation}
	The formal normal equations associated with these two instantaneous least-squares objectives are
	\begin{equation}
		X^*(n)X(n)W=X^*(n)l(n),
		\label{eq:eql}
	\end{equation}
	and
	\begin{equation}
		X^*(n)X(n)W=X^*(n)r(n)=X^*(n)\bigl(l(n)+s(n)\bigr),
		\label{eq:eqr}
	\end{equation}
	respectively. Thus, \eqref{eq:eql}--\eqref{eq:eqr} describe a family of conditions, one for each time $n$, rather than a single time-invariant optimization problem. The uniqueness of the solution to an instantaneous problem is determined by the rank of $X^*(n)X(n)$. In the present single-sample model, $X(n)\in\mathbb{C}^{1\times N}$ and therefore $\operatorname{rank}(X^*(n)X(n))\leq 1$; hence, when $N>1$, the instantaneous minimizer is generally non-unique. These conditions are used here to expose the difference between the ideal and observable instantaneous objectives. The two conditions coincide at time $n$ only if $X^*(n)s(n)=0$. Assumption~\eqref{eq:statistical-assumptions} imposes only the ensemble-level condition $\mathbb{E}\{X^*(n)s(n)\}=\bm{0}$; it does not imply $X^*(n)s(n)=0$ for each realization or at each time instant. Therefore, even when Assumption~\eqref{eq:statistical-assumptions} is satisfied, this sample-wise orthogonality condition should not be expected for an arbitrary instantaneous problem $\mathcal{P}_n$.
	
	\subsection{Iterative solution of the optimization sequence}
	LMS and related methods generally do not iterate each instantaneous problem to convergence. This is because the observable instantaneous problem $\mathcal P^{\mathrm{rx}}_n$ differs from the ideal problem $\mathcal P^{\mathrm{SI}}_n$ through the desired-signal term. Iterating toward convergence of $\mathcal P^{\mathrm{rx}}_n$ may produce a better fit to the received signal while producing a worse approximation of the SI component, especially at high SIR. Instead, these methods process the sequence $\{\mathcal P^{\mathrm{rx}}_n\}$ recursively. At time $n$, the initial coefficient vector is the value obtained after processing the preceding problem, namely $W(n)$. The current residual is
	\begin{equation}
		e(n)=r(n)-X(n)W(n),
	\end{equation}
	and LMS takes one stochastic-gradient step for the current objective:
	\begin{equation}
		W(n+1)=W(n)+\mu e(n)X^*(n),
		\label{eq:lms}
	\end{equation}
	where $\mu$ is the step size. Thus, $W(n+1)$ is both the updated coefficient vector for the current problem and the warm start for the next problem $\mathcal P^{\mathrm{rx}}_{n+1}$. NLMS, variable-step-size LMS, CG, and RLS follow the same sequential interpretation but use different normalizations, step-size rules, or search directions \cite{NLMS1,NLMS2,VSS-LMS1,VSS-LMS2,VSS-LMS3,CG,RLS}.
	
	For nonlinear SI cancellation, the same sequential formulation is used after replacing $X(n)$ with a nonlinear feature vector. For example, a memory-polynomial canceller is
	\begin{equation}
		\widetilde d(n)=\sum_{i=1}^{N}\sum_{k=0}^{p}W_{i,k}|x(n-N+i)|^{2k}x(n-N+i),
		\label{eq:d(n)nonlinear}
	\end{equation}
	and its coefficients are updated recursively from the current residual. The desired signal therefore perturbs the update of both linear and nonlinear iterative cancellers, especially when SIR is high. This motivates the operator-learning formulation in Section~\ref{sec:method}.
	
	\section{Self-Interference Cancellation Method Based on Transformer Neural Operators}
	\label{sec:method}
	This section develops the proposed operator-learning formulation for digital SI cancellation. The main contribution is the reformulation of the estimation problem: instead of updating cancellation coefficients directly from a desired-signal-contaminated instantaneous error, the method learns the temporal evolution of the SI channel from calibration data and uses this learned evolution to reconstruct the current SI. A Transformer neural operator is employed as a realization of this operator-learning framework. The following subsections present the formulation, the network implementation, the training and application procedures, and a stability analysis.
	
	\subsection{Reformulation of self-interference cancellation problem}
	\label{sec:Reformulation}
	The SI signal is determined by the local transmit signal and the SI channel, whereas the desired signal is generated by the remote transmitter. This separation motivates an operator-learning formulation: if the temporal evolution of the SI channel can be characterized, the current SI can be reconstructed without using an instantaneous coefficient update driven by the desired-signal-contaminated residual.
	
	Let $h_n$ denote the SI-channel mapping at time $n$. Exploiting temporal dependence in channel evolution, we introduce an operator $\bar{G}$ that maps a sequence of historical channel mappings to the current channel mapping:
	\begin{equation}
		\bar{G}\left(h_{n-1},h_{n-2},\ldots,h_{n-M}\right)
		= h_n,
		\label{eq:G}
	\end{equation}
	where $M$ is the number of preceding channel realizations used to characterize the channel dynamics. Thus, $\bar{G}$ represents the temporal evolution law of the SI channel in a function space.
	
	Applying the predicted current channel mapping to the known transmit vector gives the operator-learning representation of SI:
	\begin{equation}
		l(n)
		=
		\bar{G}
		\!\left(
		h_{n-1},
		h_{n-2},
		\ldots,
		h_{n-M}
		\right)
		\bigl(\bm{x}_n\bigr),
		\label{eq:operator_si}
	\end{equation}
	where
	\(
	\bar{G}(h_{n-1},h_{n-2},\ldots,h_{n-M})
	\)
	represents the channel mapping at the current time instant. Learning $\bar{G}$ therefore reconstructs SI from historical SI-channel behavior and the known local transmit signal. Unlike few-shot calibration methods that treat context samples from a stationary calibration interval as exchangeable \cite{Balaji2026}, $\bar{G}$ is an ordered temporal mapping from preceding channel realizations to the current channel realization. This is the central distinction between the proposed framework and conventional error-driven iterative cancellation.

	Let $G_\theta$ denote a Transformer neural operator with parameters $\theta$. It is used to approximate $\bar{G}$:
	\begin{equation}
		\theta^*=\sideset{}{}{\mathop{\min}}_{\theta \in \Theta } \parallel \bar{G}-G_\theta\parallel, \label{eq:Gt}
	\end{equation}
	where $\Theta$ is the parameter space. For a signal segment of length $T$, the empirical training loss is
	\begin{equation}
		J(\theta)=\frac{1}{T}\sum_{n=1}^{T} \mid l(n)-d_\theta(n)|^2.
	\end{equation}
	where $d_\theta(n)$ is the network output:
	\begin{equation}
		d_\theta(n)=G_{\theta}\left(h_{n-1},h_{n-2},\ldots,h_{n-M}\right)(\bm{x}_n).
	\end{equation}
	
	\subsection{Transformer Neural Operators}
	\label{sec:transformer}
	We use a Transformer neural operator based on self-attention. It maps the input function space $A(D;\mathbb{R}^{d_a})$ to the output function space $U(D;\mathbb{R}^{d_u})$:
	$G:A(D;\mathbb{R}^{da})\times \Theta \longrightarrow U(D;\mathbb{R}^{du}).$
	The target operator is $G^+:A(D;\mathbb{R}^{d_a})\longrightarrow U(D;\mathbb{R}^{d_u})$. The neural-operator realization has the form
	\begin{align}
		G(a,\theta)=(T_{out}\circ E_L\circ T_{in})(a,\theta),
		\label{eq:G_in}
	\end{align}
	for $a\in A(D;\mathbb{R}^{d_a})$ and $\theta\in\Theta$.
	
\begin{figure}[H]
		\centering 
		\includegraphics[width=1.0\textwidth]{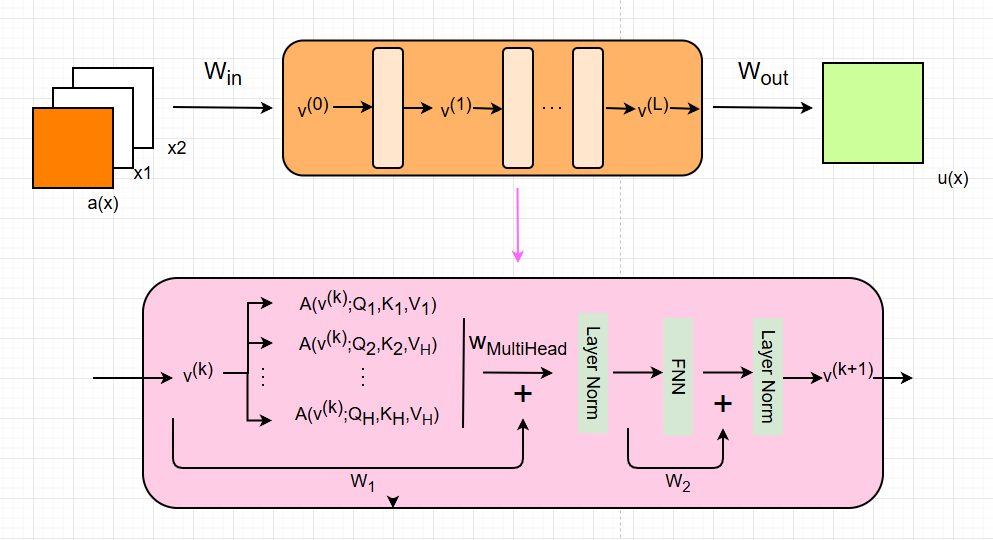}
		\caption{Transformer Neural Operator}
		\label{fig:ima1} 
	\end{figure}

	In \eqref{eq:G_in}, $T_{in}:A(D;\mathbb{R}^{d_a})\times\Theta\rightarrow V(D;\mathbb{R}^{d_v})$ and $T_{out}:V(D;\mathbb{R}^{d_v})\times\Theta\rightarrow U(D;\mathbb{R}^{d_u})$ are input and output lifting operators on an embedded function space $V$. The operator $E_L:V(D;\mathbb{R}^{d_v})\times\Theta\rightarrow V(D;\mathbb{R}^{d_v})$ is the neural-operator analogue of an $L$-layer Transformer encoder \cite{transformer}. For brevity, write $T_{in}(a)=T_{in}(a,\theta)$ and $T_{out}(a)=T_{out}(a,\theta)$. With $v^0=T_{in}(a)$, each encoder layer is
	\begin{align}
		v^{l-1} &= W_1v^{l-1} + A_{MultiHead}v^{l-1}, \\
		v^{l-1} &= F_{layernorm}(v^{l-1}), \\
		v^{l-1} &= W_2v^{l-1} + F_{NN}(v^{l-1}), \\
		v^{l}&=F_{layernorm}(v^{l-1}),
	\end{align}
	for $l=1,2,\ldots,L$, with $E_L(v^0)=v^L$. The operators $W_1$, $W_2$, $A_{MultiHead}$, $F_{layernorm}$, and $F_{NN}$ have independent learnable parameters. The multi-head self-attention operator $A_{MultiHead}:V\rightarrow V$ is
	\begin{equation}
		\begin{split}
			(A_{MultiHead}(v))(x)=W_{MultiHead}(A(v,\theta_1)(x),\\...,A(v,\theta_H)(x)),
		\end{split}
	\end{equation}
	for $v\in V$, where $H$ is the number of attention heads and $W_{MultiHead}:V\rightarrow V$ is a learned linear transformation. The single-head attention map $A$ is
	\begin{equation}
		\begin{split}
			A(v,\theta_i)(x)=\frac{\int_{D}exp(<Q_iv(x),K_iv(y)>)V_iv(y)dy}{\int_{D}exp(<Q_iv(x),K_iv(s)>)ds}\\ \quad\forall x\in D, i=1,...,H.
		\end{split}
	\end{equation}
	Here, $\langle\cdot,\cdot\rangle$ denotes the Euclidean inner product, and $Q$, $K$, and $V$ are learned linear transformations. The maps $W_1,W_2:V\rightarrow V$ are linear operators, $F_{NN}$ is a single-hidden-layer fully connected network, and $F_{layernorm}:V\rightarrow V$ is layer normalization:
	\begin{equation}
		(F_{layernorm}(v;\gamma,\beta)(x))_k=\gamma_k\cdot\frac{(v(x))_k-m(v(x))}{\sqrt{\sigma^2(v(x))+\varepsilon}}+\beta_k,
	\end{equation}
	for $k=1,...,dv$, $\forall v\in V$, and $x\in D\subset R^d$. Here, $(\cdot)_k$ refers to the $k$-th component of a vector, $\varepsilon\in R^+$ is a fixed parameter, and $\gamma_k,\beta_k\in R$ are learnable parameters. The definitions of $m$, $\sigma$ are as follows:
	\begin{equation}
		\begin{aligned}
			m(y)&=\frac{1}{dv}\cdot\parallel  y\parallel_1,\\
			\sigma^2(y)&=\frac{1}{dv}\sum_{k=1}^{dv} (y_k-m(y))^2,
		\end{aligned}
	\end{equation}
	for any $y\in R^{dv}$. The entire Transformer Neural Operator can be illustrated in Fig. \ref{fig:ima1},
	where $W_{in}$ and $W_{out}$ respectively denote the operator $T_{in}(a(x))$ and $T_{out}(v^{L})$.
	
	\subsection{Method setting}
	\label{sec:Method setting}
	According to \eqref{eq:G}, the input of $\bar{G}$ is $h_i$, $i=n-M,\ldots,n-1$, and its output is $h_n$.
	Information about $h_i$ is contained in the input--output relation between $X(i)$ and $l(i)$, while $X(n)$ is the argument of the current mapping $h_n$. Hence, the input of $\bar{G}$ is represented by $(X(n),X(n-1),\ldots,X(n-M),l(n-1),\ldots,l(n-M))$. Since $X(n)=(x(n-N+1),\ldots,x(n))$, the input of $G_\theta$ can be written as $(x(n),\ldots,x(n-M-N+1),l(n-1),\ldots,l(n-M))$. The ordering of these historical observations is retained to characterize temporal SI-channel evolution rather than only a stationary calibration context.
	
	The neural operator is trained offline in a desired-signal-free setting, where $r(n)=l(n)$ and the SI labels are available. During online operation, $r(n)$ may contain the desired signal and $l(n)$ is unavailable. The trained operator $G_{\theta^*}$ therefore uses $(x(n),\ldots,x(n-M-N+1),r(n-1),\ldots,r(n-M))$ as its input. Algorithms~\ref{alg1} and \ref{alg:alg2} summarize the procedures for static and time-varying channels, respectively.
	
	\begin{algorithm}[h]
		\caption{Static-channel scenario.}\label{alg:alg1}
		\begin{algorithmic}
			\STATE 
			\STATE {\textsc{Offline Training:}}
			\STATE \hspace{0.5cm}$ \textbf{Step 1: }$ Use 3,000 samples for training. The input is $(x(n),\ldots,x(n-M-N+1),l(n-1),\ldots,l(n-M))$, and the target is $l(n)$.
			\STATE \hspace{0.5cm}$ \textbf{Step 2: }$ Train for 60,000 iterations using the MSE loss and the Adam optimizer.
			\STATE \hspace{0.5cm}$ \textbf{Step 3: }$ Save the trained model as $G_{\theta^*}$.
			
			\STATE 
			\STATE {\textsc{Online Test:}}
			\STATE \hspace{0.5cm}$ \textbf{Step 4: }$ Apply $G_{\theta^*}$ to $(x(n),\ldots,x(n-M-N+1),r(n-1),\ldots,r(n-M))$.
			\STATE \hspace{0.5cm}\textbf{return}  $d(n)$
		\end{algorithmic}
		\label{alg1}
	\end{algorithm}
	
	The difference between the static and dynamic settings is whether the SI channel changes over time. In the static setting, the offline-trained operator can be applied directly online. In the dynamic setting, channel variation can create a mismatch between the training and test distributions and may reduce prediction accuracy.
	
	To track rapidly varying channels, the model is periodically updated using newly collected observations. The update interval depends on the severity of the channel variation: faster dynamics generally require more frequent updates. The dynamic data set used here is deliberately challenging: the SI channel undergoes an abrupt change approximately every 500 samples, while it varies gradually within each 500-sample interval. The operator is retrained after every 20,000 received samples. Specifically, a pilot-like, desired-signal-free calibration interval is used to acquire 1,000 newly labelled SI samples $l(n)$, and the operator is fine-tuned using these samples, as summarized in Algorithm~\ref{alg:alg2}. This procedure targets continuous tracking under nonstationary channel evolution, rather than few-shot prediction within a single quasi-static calibration interval.
	
	\begin{algorithm}[h]
		\caption{Time-varying-channel scenario.}\label{alg:alg2}
		\begin{algorithmic}
			\STATE 
			\STATE {\textsc{Offline Training:}}
			\STATE \hspace{0.5cm}$ \textbf{Step 1: }$ 
			Same as Step 1 of Algorithm~\ref{alg1}.
			\STATE \hspace{0.5cm}$ \textbf{Step 2: }$ Train for 100,000 iterations using the MSE loss and the Adam optimizer.
			\STATE \hspace{0.5cm}$ \textbf{Step 3: }$ Save the trained model as $G_{\theta^*}$.
			
			\STATE 
			\STATE {\textsc{Online Test:}}
			\STATE \hspace{0.5cm}$ \textbf{Step 4: }$ Load the offline-trained model $G_{\theta^*}$.
			\STATE \hspace{0.5cm}$ \textbf{Step 5: }$ At a pilot-like, desired-signal-free calibration interval, acquire 1,000 labelled SI samples $l(n)$ and fine-tune the model for 600 iterations using the settings in Steps 1 and 2.
			\STATE \hspace{0.5cm}$ \textbf{Step 6: }$ Use the updated model to predict the following 20,000 samples with input $(x(n),\ldots,x(n-M-N+1),r(n-1),\ldots,r(n-M))$.
			\STATE \hspace{0.5cm}\textbf{return}  $d(n)$ 
			\STATE \hspace{0.5cm}$ \textbf{Step 7: }$ Use the next 1,000 samples and return to Step 5.
		\end{algorithmic}
	\end{algorithm}

	
	\subsection{Analysis of input-perturbation sensitivity to desired-signal contamination}
	\label{sec:Analysis}
	The proposed framework is trained with inputs that do not contain the desired signal. Its online input is formed by replacing the unobservable historical SI samples with received samples, which include the desired signal. The following result characterizes the sensitivity of the learned operator to this replacement. Let $T_0=(x(n),\ldots,x(n-M-N+1),l(n-1),\ldots,l(n-M))$ be the clean input and $T_1=(x(n),\ldots,x(n-M-N+1),r(n-1),\ldots,r(n-M))$ be the online input. For a given transmit sequence, define $\|\cdot\|_x=\|\cdot\|/\operatorname{mean}\{|x(n)|\}$.
	Assume that $G_{\theta^*}$ is continuous: for every $\varepsilon>0$, there exists $\delta(\varepsilon)>0$ such that $\|T_0-T_1\|_x<\delta(\varepsilon)$ implies $|G_{\theta^*}(T_0)-G_{\theta^*}(T_1)|<\varepsilon$. We then obtain the following input-perturbation sensitivity result.
	
	{\textbf{Theorem 1}} (Sensitivity to desired-signal contamination):
	Assume that $G_{\theta^*}$ is continuous. For every $\varepsilon>0$, there exists $\delta(\varepsilon)>0$ such that, if the desired signal $s(n)$ and transmit signal $x(n)$ satisfy
	\begin{equation}
		\frac{max\parallel s(n)\parallel}{mean\parallel x(n)\parallel}< \frac{\delta(\varepsilon)}{M},  \label{cond}
	\end{equation}
	then
	\begin{equation}
		\mid G_{\theta^*}(T_0)-G_{\theta^*}(T_1)\mid< \varepsilon . \label{conc}
	\end{equation}
	
	\emph{Proof:} Direct calculation gives
	\begin{equation}
		\begin{aligned}
			\parallel T_0-T_1\parallel_x&=\parallel(0,...,0,s(n-1),...s(n-M))\parallel_x\\
			&\le M\frac{max\parallel s(n)\parallel}{mean\parallel x(n)\parallel}\\
			&<\delta(\varepsilon).
		\end{aligned}
	\end{equation}
	Continuity of $G_{\theta^*}$ yields \eqref{conc}. \hfill $\square$
	
	In practical full-duplex systems, the desired signal originates from the remote transmitter. It may exceed the residual SI level, yet it is typically much weaker than the local transmit signal; thus, condition~\eqref{cond} is meaningful in the intended operating regime, including the high-SIR cases considered in the experiments. The proof characterizes the sensitivity to the replacement of clean historical SI samples by received samples under condition~\eqref{cond}; it neither establishes convergence nor guarantees performance for an arbitrary SIR. High SIR concerns the ratio between the desired signal and residual SI, whereas condition~\eqref{cond} controls the desired signal relative to the local transmit signal. The proof therefore provides a sufficient condition under which high-SIR operation may remain stable, while the high-SIR cancellation gains are numerical findings. This property follows from the continuity of the learned operator and the proposed clean-training/contaminated-inference input structure, rather than from a particular attention mechanism alone.
	
	\section{Numerical Experiments
	}
	\label{sec:Numerical Experiments}
	We evaluate the proposed method under different desired-signal contamination levels to assess its cancellation performance and robustness. The desired-signal-to-SI ratio (SIR) and normalized mean-squared error (NMSE), both in dB, are defined as
	\begin{equation}
		\begin{aligned}
			\mathrm{SIR}&=20\log_{10}\frac{\lVert \bm{s}\rVert_2}{\lVert \bm{l}\rVert_2},\\
			\mathrm{NMSE}&=20\log_{10}\frac{\lVert \bm{err}\rVert_2}{\lVert \bm{l}\rVert_2}.
		\end{aligned}
	\end{equation}
	Here, $\bm{s}$ is the vector formed by $s(n)$ over the evaluation interval; $\bm{l}$ and $\bm{err}$ are defined analogously for the SI signal $l(n)$ and cancellation error $err(n)$, respectively. As discussed in Section~\ref{sec:model}, the desired signal $s(n)$ contaminates SI estimation. Therefore, the desired-signal-free scenario considered below means $s(n)=0$; it is not a 0-dB-SIR scenario.
	
	\subsection{Static-channel results}
	The static scenario considers a time-invariant SI channel. In this experiment, the channel vector is $(0,0,\ldots,0.001)$, so that $l(n)=0.001x(n)$. The data are synthetically generated and processed using a continuous sliding window. The first 3,000 samples are used for offline training, and the trained network is evaluated on disjoint data segments of different lengths.
	
	We use a Transformer neural operator with ReLU activation and the Adam optimizer. Training is performed for 60,000 iterations. Because the network operates on real-valued inputs, each complex sample is represented by its real and imaginary parts; the output similarly represents the real and imaginary parts of $d(n)$. The input dimension is $2(2M+N)$ and the output dimension is two. The hyperparameters are listed in Table~\ref{tab:parameters}, and the static-channel procedure is given in Algorithm~\ref{alg:alg1}.
	
	\begin{table}[H]
		\centering
		\caption{Hyperparameters for a static scene}
		\begin{tabular}{ccc}
			\hline
			Parameters&Value &Explanation\\ 
			\hline
			$emb$     &256&the number of neurons in each layer of $E_L$\\ 
			\hline
			$hidden$ &512& the hidden layer size of the FNN \\ 
			\hline
			$heads$ &2&the number of attention heads \\
			\hline
			$layers$&1  &the number of layers in $E_L$ \\
			\hline
			$N$&10 &the tap coefficient\\
			\hline
			$M$&10  &the truncation number\\
			\hline
		\end{tabular}
		\label{tab:parameters}
	\end{table}
	
	\begin{itemize}
		\item \texttt{number of iterations:} 60,000
		\item \texttt{learning rate:} iterations $0$--$20{,}000$, $2\times10^{-4}$; iterations $20{,}000$--$40{,}000$, $4\times10^{-5}$; iterations $40{,}000$--$60{,}000$, $8\times10^{-6}$.
	\end{itemize}
	
	We evaluate NMSE under different desired-signal contamination levels and over evaluation segments of different lengths. The results are reported in Table~\ref{tab:results}. For compactness, `Free' denotes the desired-signal-free scenario.
	
	\begin{table}[H]
		\centering
		\caption{Results of NMSE in static scenario}
		\begin{tabular}{ccccc}
			\hline
			data&Free& -10db   & 10db   &30db \\ 
			\hline
			training set & -36.75dB  &-36.75dB  &-36.75dB  &-36.75dB \\ 
			\hline
			3000 test set  & -16.35dB  &-16.35dB  &-16.35dB  &-16.30dB \\ 
			\hline
			30,000 test set  & -14.76dB  &-14.76dB  &-14.76dB  &-14.72dB \\
			\hline
			100,000 test set & -13.51dB  &-13.51dB  &-13.51dB  &-13.48dB\\
			\hline
			300,000 test set & -15.16dB  &-15.16dB  &-15.16dB  &-15.13dB\\
			\hline
		\end{tabular}
		\label{tab:results}
	\end{table}
	Table~\ref{tab:results} shows that varying the SIR has little effect on the proposed method in this static experiment. The largest displayed difference between the desired-signal-free and 30-dB-SIR cases is no greater than 0.05 dB. Thus, the proposed digital-domain SI canceller achieves performance comparable to that in the desired-signal-free case. This numerical observation is consistent with the input-perturbation sensitivity analysis in Section~\ref{sec:Analysis} under condition~\eqref{cond}.

	\subsection{Time-varying-channel results}
	We next consider a dynamic SI channel. The data set is deliberately challenging: the channel undergoes an abrupt change approximately every 500 samples and varies gradually within each 500-sample interval. An initial network is trained using 3,000 desired-signal-free samples and then saved. After every 20,000 received samples, a pilot-like desired-signal-free calibration interval provides 1,000 labelled SI samples for 600 fine-tuning iterations. The updated operator is then used to predict the subsequent 20,000 samples. This calibration--prediction cycle is repeated over 12 million samples, following Algorithm~\ref{alg:alg2}. The experiment therefore evaluates continuous SI tracking under nonstationary channel evolution, rather than few-shot prediction within a stationary calibration interval.
	
	For the initial network trained with 3,000 samples, we compare different hyperparameter configurations. For model selection, two disjoint held-out segments of 3,000 and 30,000 samples are used as validation data, while a 60,000-sample held-out segment is retained for final testing. Table~\ref{tab:parameters2} lists the configurations, and Table~\ref{tab:result2} reports the corresponding results.
	
	\begin{table}[H]
		\centering
		\caption{Different hyperparameters of network}
		\begin{tabular}{ccccccc}
			\hline
			Model&$emb$ & $hidden$& $heads$&$layers$&$N$&$M$ \\ 
			\hline
			M1 &64 &128&4&4&10&10\\ 
			\hline
			M2 &128 &256&4 &4&10&10\\ 
			\hline
			M3 &128 &256&8 &5&10&10\\
			\hline
			M4 &32 & 64 &4 &3&10&10\\
			\hline
			M5 &128 & 256&4&2&10&10\\
			\hline
			M6 &128 & 256&4&1&10&10\\
			\hline
			M7 &128 & 256&4&1&20&10\\
			\hline
			M8 &128 & 256&4&1&30&10\\
			\hline
			M9 &128 & 256&4&1&40&10\\
			\hline
		\end{tabular}
		\label{tab:parameters2}
	\end{table}
	\begin{itemize}
		\item \texttt{number of iterations:} 100,000
		\item \texttt{learning rate:} iterations $0$--$20{,}000$, $2\times10^{-4}$; $20{,}000$--$40{,}000$, $4\times10^{-5}$; $40{,}000$--$60{,}000$, $8\times10^{-6}$; $60{,}000$--$80{,}000$, $2\times10^{-6}$; $80{,}000$--$100{,}000$, $4\times10^{-7}$.
	\end{itemize}

	\begin{table}[H]
		\centering
		\caption{Results of NMSE under different hyperparameters}
		\begin{tabular}{ccccc}
			\hline
			Model & Training & Validation (3k) & Validation (30k) & Test (60k)\\
			\hline
			M1 & -15.05dB & -14.09dB & -13.08dB & -10.90dB\\
			M2 & -16.31dB & -15.00dB & -12.81dB & -11.09dB\\
			M3 & -15.10dB & -14.51dB & -12.70dB & -10.70dB\\
			M4 & -13.40dB & -11.89dB & -11.70dB & -10.18dB\\
			M5 & \textbf{-18.51}dB & -15.10dB & -14.10dB & \textbf{-11.85}dB\\
			M6 & -15.56dB & -14.32dB & -13.45dB & -11.17dB\\
			M7 & -17.24dB & -15.80dB & -14.31dB & -11.62dB\\
			M8 & -18.25dB & \textbf{-16.15}dB & \textbf{-14.58}dB & -11.65dB\\
			M9 & -17.26dB & -15.40dB & -13.75dB & -11.30dB\\
			\hline
		\end{tabular}
		\label{tab:result2}
	\end{table}

	As shown in Table~\ref{tab:result2}, M8 achieves the lowest NMSE on the 3k and 30k validation segments and is therefore selected as the initial model for the remaining dynamic experiments. The 60k test segment is not used for this selection. During the 12-million-sample evaluation, the operator is fine-tuned after every 20,000 received samples using 1,000 labelled SI samples from the pilot-like calibration interval for 600 iterations. The learning rate is $8\times10^{-6}$ for iterations 1--400 and $4\times10^{-6}$ for iterations 401--600. The procedure is specified in Algorithm~\ref{alg:alg2}. This update schedule is evaluated under the highly dynamic channel setting described above.
	
\begin{figure}[H]
		\centering
		\begin{minipage}[b]{0.48\textwidth}
			\centering
			\includegraphics[width=\textwidth]{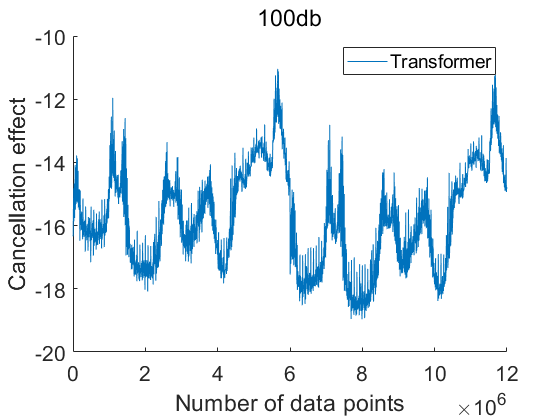}
			\caption{NMSE of 100 dB SIR }
			\label{fig:image10}
		\end{minipage}
		\hfill
		\begin{minipage}[b]{0.48\textwidth}
			\centering
			\includegraphics[width=\textwidth]{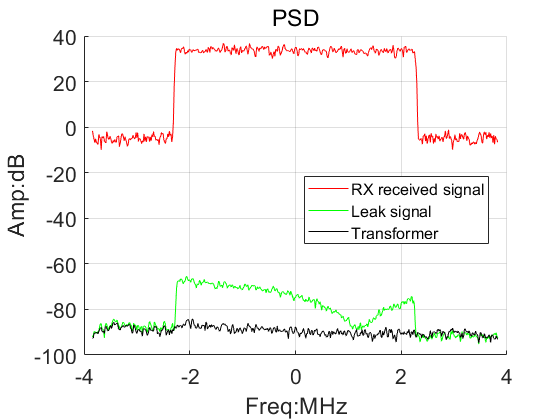}
			\caption{PSD of 100 dB SIR }
			\label{fig:image11}
		\end{minipage}
	\end{figure}

	We compare the proposed Transformer-neural-operator-based digital SI canceller with LMS, NLMS\cite{NLMS1,NLMS2}, CG\cite{CG}, RBFNN\cite{NN-base1}, and nonlinear LMS (NL-LMS) \cite{NONLMS1,NONLMS2} under different SIR conditions. RBFNN is not included in the desired-signal-free scenario because it requires prior estimation of the probability density function (PDF) of the desired signal. In the legends, $b$ and $\mu$ denote step-size parameters, whereas $d$ denotes the CG weight parameter. For each baseline, an approximate parameter range is first identified and then searched on an exponential base-2 grid; the best-performing parameter in this search is reported. For example, the RBFNN parameter $b$ is searched from $2^{-50}$ to $2^{-40}$. The baseline grid search is performed retrospectively over the complete experimental trajectory, so the displayed baseline curves represent best-case offline-tuned performance rather than an online parameter-selection procedure. In contrast, the proposed method uses the validation segments described above for model selection. This setting is favorable to the baselines; therefore, any observed advantage of the proposed method is conservative.
	\\
	\indent (1) Desired-signal-free scenario
	\begin{figure}[H]
		\centering
		\begin{minipage}[t]{0.48\textwidth}
			\centering
			\includegraphics[width=\textwidth]{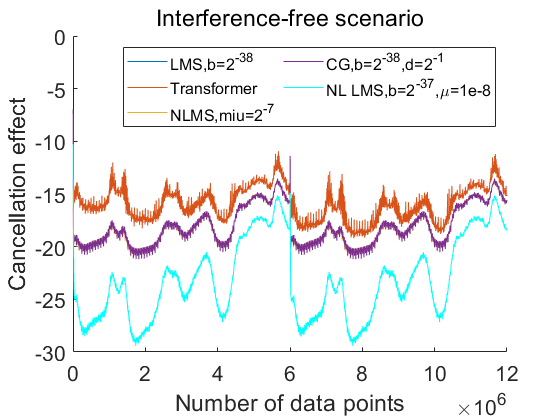}
			\caption{Desired-signal-free NMSE}
			\label{fig:image2}
		\end{minipage}
		\hfill
		\begin{minipage}[t]{0.48\textwidth}
			\centering
			\includegraphics[width=\textwidth]{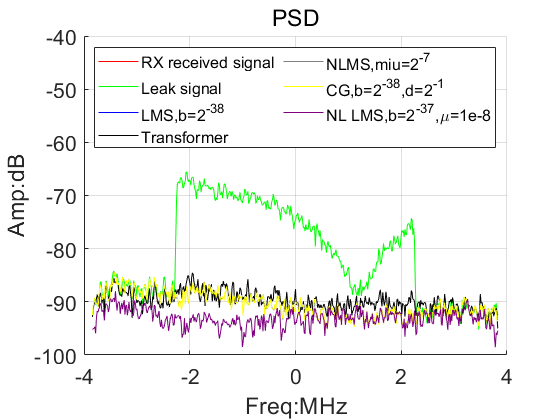}
			\caption{Desired-signal-free PSD}
			\label{fig:image3}
		\end{minipage}
	\end{figure}
	Figures~\ref{fig:image2}--\ref{fig:image3} show that, in the desired-signal-free static scenario, the proposed method is initially slightly inferior to the LMS-family methods. The periodic pilot-assisted updates considered in the subsequent dynamic-channel experiment improve the operator prediction as the model is adapted to new channel observations.
	\\
	\indent (2) $-10$ dB SIR scenario
	\begin{figure}[H]
		\centering
		\begin{minipage}[b]{0.48\textwidth}
			\centering
			\includegraphics[width=\textwidth]{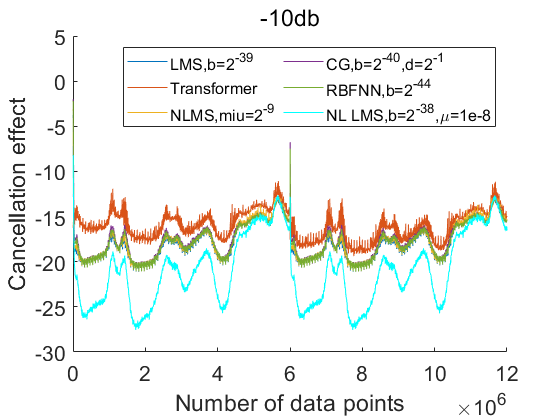}
			\caption{NMSE of -10 dB SIR}
			\label{fig:image4}
		\end{minipage}
		\hfill
		\begin{minipage}[b]{0.48\textwidth}
			\centering
			\includegraphics[width=\textwidth]{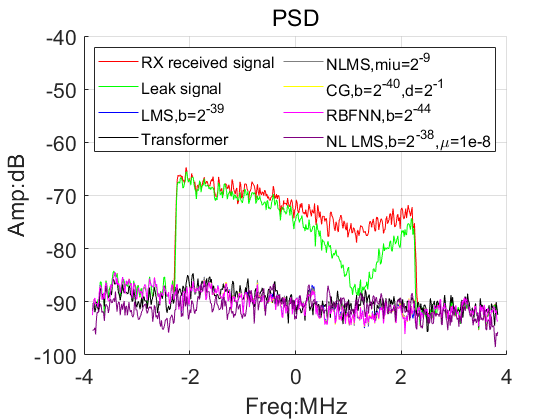}
			\caption{PSD of -10 dB SIR }
			\label{fig:image5}
		\end{minipage}
	\end{figure}
	As shown in Figs.~\ref{fig:image4}--\ref{fig:image5}, at $-10$ dB SIR the LMS-family methods degrade slightly, whereas the proposed method changes little in this experiment.
	\indent (3) 10 dB SIR scenario
	\begin{figure}[H]
		\centering
		\begin{minipage}[b]{0.48\textwidth}
			\centering
			\includegraphics[width=\textwidth]{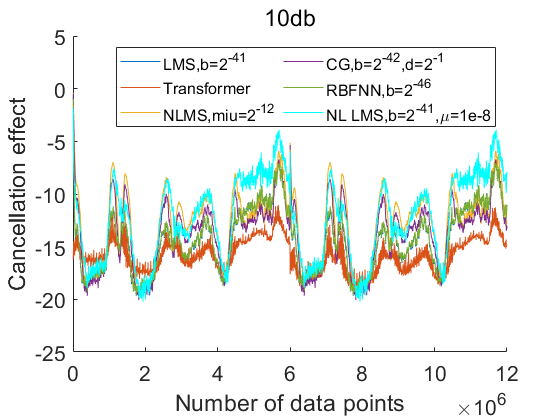}
			\caption{NMSE of 10 dB SIR}
			\label{fig:image6}
		\end{minipage}
		\hfill
		\begin{minipage}[b]{0.48\textwidth}
			\centering
			\includegraphics[width=\textwidth]{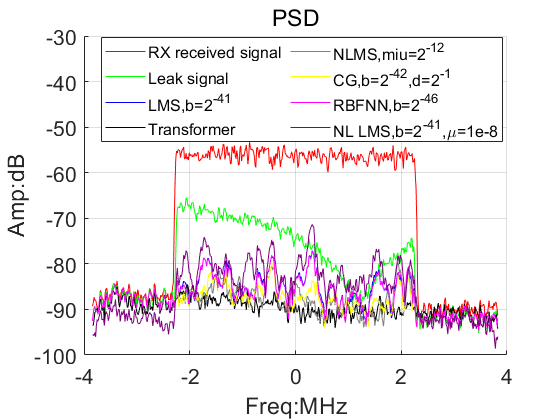}
			\caption{PSD of 10 dB SIR }
			\label{fig:image7}
		\end{minipage}
	\end{figure}
	Figures~\ref{fig:image6}--\ref{fig:image7} indicate that, as SIR increases to 10 dB, the performance of LMS further deteriorates and NLMS exhibits a similar but more pronounced degradation. In contrast, the proposed method maintains stable performance and shows less oscillatory behavior.
	\\
	\indent (4) 30 dB SIR scenario
	\begin{figure}[H]
		\centering
		\begin{minipage}[b]{0.48\textwidth}
			\centering
			\includegraphics[width=\textwidth]{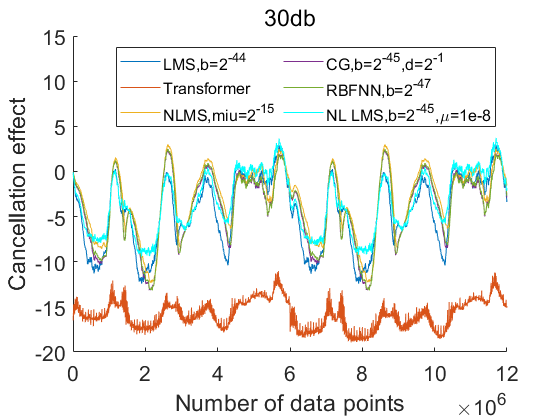}
			\caption{NMSE of 30 dB SIR}
			\label{fig:image8}
		\end{minipage}
		\hfill
		\begin{minipage}[b]{0.48\textwidth}
			\centering
			\includegraphics[width=\textwidth]{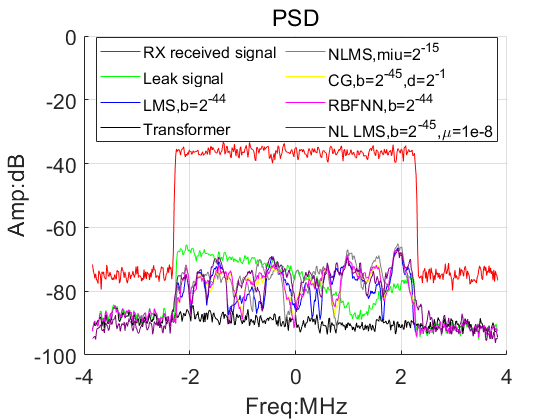}
			\caption{PSD of 30 dB SIR}
			\label{fig:image9}
		\end{minipage}
	\end{figure}
	Figures~\ref{fig:image8}--\ref{fig:image9} illustrate the results at 30 dB SIR. The iterative baseline methods exhibit opposite cancellation, meaning that their residual power after cancellation exceeds the uncancelled SI power. In contrast, the proposed method remains stable and improves NMSE by 5--15 dB relative to the considered baseline methods.\\
	\indent (5) 100 dB SIR scenario\\
	\indent We also consider an extreme 100-dB-SIR case to assess whether the proposed method remains robust to severe desired-signal contamination. The results are shown in Figs.~\ref{fig:image10}--\ref{fig:image11}.
	Because all compared baseline algorithms exhibit opposite cancellation over the displayed interval, their curves are omitted from Figs.~\ref{fig:image10}--\ref{fig:image11}. Even under this high-SIR condition, the proposed method maintains performance comparable to that in the desired-signal-free scenario.
	
	\subsection{Results with Independently Time-Varying SI Channel and Desired-Signal Strength}
	In this experiment, the SI channel and desired-signal strength vary independently over time. The desired-signal condition switches among four values (desired-signal-free, $-10$ dB, 10 dB, and 30 dB SIR) every 20,000 samples. We evaluate whether the proposed method retains stable cancellation performance under these independently time-varying conditions. Model M8 is used for initialization. The data selection, input representation, number of training iterations, and learning-rate schedule for periodic updates are the same as in the preceding dynamic-channel experiment.
	
\begin{figure}[H]
		\centering
		\begin{minipage}[t]{0.48\textwidth}
			\centering
			\includegraphics[width=\textwidth]{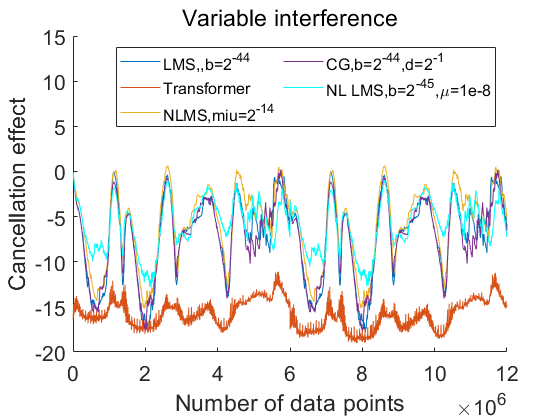}
			\caption{NMSE of varying SIR}
			\label{fig:image12}
		\end{minipage}
		\hfill
		\begin{minipage}[t]{0.48\textwidth}
			\centering
			\includegraphics[width=\textwidth]{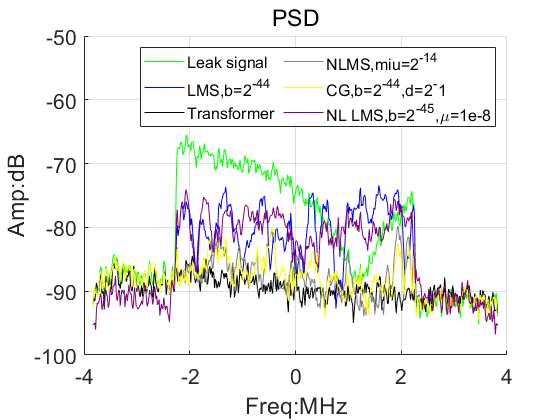}
			\caption{PSD of varying SIR}
			\label{fig:image13}
		\end{minipage}
	\end{figure}

	RBFNN is omitted because it requires prior estimation of the desired-signal PDF, which changes rapidly in this setting. The results are shown in Figs.~\ref{fig:image12}--\ref{fig:image13}.
	In this scenario, LMS-type methods achieve their best high-SIR performance only when their parameters are tuned for the strongest desired-signal contamination. However, this configuration impairs channel tracking at lower SIR, degrading overall performance when the SI channel and desired-signal strength vary independently over time. In contrast, the proposed method maintains stable performance in this independently time-varying environment.
	
	\section{Conclusion}
	\label{sec:conclusion}
	This paper presented a digital-domain residual-SI cancellation method for in-band full-duplex systems based on Transformer neural operators. The method operates after passive suppression and analog cancellation. Its main idea is to reformulate digital-domain SI cancellation as an operator-learning problem: the temporal evolution of the residual SI channel is learned from historical observations and then used to reconstruct the current residual SI. This formulation reduces the reliance on the desired-signal-contaminated instantaneous updates used by conventional iterative cancellers. Theorem~1 characterizes the input-perturbation sensitivity of the resulting input--output structure to desired-signal contamination under the stated desired-to-local-transmit-signal condition.
	
	In the reported experiments, the proposed method performs comparably to conventional LMS-type methods at low SIR and remains stable as SIR increases. In high-SIR settings, including the 100-dB case, the method maintains performance close to that observed in the desired-signal-free setting, whereas the considered iterative methods deteriorate. The dynamic experiments further show that the method retains this behavior when the desired-signal level and the SI channel vary independently over time.

	Future work will investigate alternative operator-learning architectures and more efficient adaptation mechanisms for rapidly time-varying SI channels.

	\section*{Data availability}
	The data supporting the findings of this study were provided by a collaborating laboratory and are not publicly available because of confidentiality restrictions.




\begin{thebibliography}{1}
		\bibliographystyle{IEEEtran}
		
		\bibitem{Full-Duplex1}
		L. Mohammadi, Z. Mobini, D. Galappaththige, and C. Tellambura, “A Comprehensive Survey on Full-Duplex Communication: Current Solutions, Future Trends, and Open Issues,” in \emph{IEEE Commun. Surveys Tuts.}, vol. 25, no. 4, pp. 2190--2244, 4th Quart. 2023.
		
		\bibitem{Full-Duplex2}
		A. Sabharwal, P. Schniter, D. Guo, D. W. Bliss, S. Rangarajan, and R. Wichman, “In-Band Full-Duplex Wireless: Challenges and Opportunities,” in \emph{IEEE J. Sel. Areas Commun.}, vol. 32, no. 9, pp. 1637--1652, Sep 2014.
		
		\bibitem{Full-Duplex3}
		B. Smida, A. Sabharwal, G. Fodor, G. C. Alexandropoulos, H. A. Suraweera, and C. -B. Chae, “Full-Duplex Wireless for 6G: Progress Brings New Opportunities and Challenges,” in \emph{IEEE J. Sel. Areas Commun.}, vol. 41, no. 9, pp. 2729--2750, Sep 2023.
		
		\bibitem{Full-Duplex4}
		K. E. Kolodziej, B. T. Perry and J. S. Herd, “In-Band Full-Duplex Technology: Techniques and Systems Survey,” in \emph{IEEE Transactions on Microwave Theory and Techniques}, vol. 67, no. 7, pp. 3025--3041, Jul 2019.
		
		\bibitem{self-interference1} 
		W. -S. Liao, O. Zhao, K. Li, H. Kawasaki, and T. Matsumura, “Implementation of In-Band Full-Duplex Communication System with Self-Interference Cancellation Using Adaptive Filter,” in \emph{Proc. IEEE VTS Asia Pac. Wireless Commun. Symp. (APWCS)}, Tainan city, Taiwan, 2023, pp. 1--5.
		
		\bibitem{self-interference2} 
		C. D. Nwankwo, L. Zhang, A. Quddus, M. A. Imran and R. Tafazolli, “A Survey of Self-Interference Management Techniques for Single Frequency Full Duplex Systems,” in \emph{IEEE Access}, vol. 6, pp. 30242--30268, 2018.
		
		\bibitem{self-interference3} 
		D. Ahmed and A. M. Eltawil, “All-Digital Self-Interference Cancellation Technique for Full-Duplex Systems,” in \emph{IEEE Transactions on Wireless Communications}, vol. 14, no. 7, pp. 3519--3532, Jul 2015.
		
		\bibitem{self-interference4} 
		H. Luo, M. Holm and T. Ratnarajah, “On the performance of active analog self-interference cancellation techniques for beyond 5G systems,” in \emph{China Communications}, vol. 18, no. 10, pp. 158--168, Oct 2021.
		
		\bibitem{multi-sic}
		D. Bharadia, E. McMilin, and S. Katti, “Full duplex radios,” in \emph{Proceedings of the ACM SIGCOMM Conference}, 2013, pp. 375--386.
		
		\bibitem{digital-sic}
		B. Yu et al., “Full duplex communication with practical self-interference cancellation implementation,” in \emph{ICC 2022-IEEE International Conference on Communications}, 2022, pp. 1100--1105.
		
		\bibitem{LMS1} 
		B. Widrow and M. E. Hoff, “Adaptive switching circuits,” in \emph{Neurocomputing: Foundations of Research}. Cambridge, MA, USA: MIT Press, 1988, pp. 123--134.
		
		\bibitem{LMS2}
		H. Lee, J. Kim, G. Choi, I. P. Roberts, J. Choi, and N. Lee, “Nonlinear Self-Interference Cancellation With Adaptive Orthonormal Polynomials for Full-Duplex Wireless Systems,” in \emph{IEEE Transactions on Wireless Communications}, vol. 24, no. 7, pp. 5796--5810, Jul 2025.
		
		\bibitem{LMS3}
		Z. Li et al., “An augmented nonlinear LMS for digital self-interference cancellation in full-duplex direct-conversion transceivers,” 
		\emph{IEEE Trans. Signal Process.}, vol. 66, no. 15, pp. 4065--4078, Aug 2018.
		
		\bibitem{NLMS1}
		C. Paleologu, J. Benesty, and S. Ciochină, “An overview on optimized NLMS algorithms for acoustic echo cancellation,” \emph{EURASIP J. Adv. Signal Process.}, vol. 2015, no. 1, pp. 1-19, Dec. 2015.
		
		\bibitem{NLMS2}
		A. Ali, M. Moinuddin and T. Y. Al-Naffouri, “NLMS is More Robust to Input-Correlation Than LMS: A Proof,” in \emph{IEEE Signal Processing Letters}, vol. 29, pp. 279-283, 2022.
		
		\bibitem{Sign-LMS}
		S. Choudhary, S. S. Pradhan, and S. N. Merchant, “A SPT treatment to the realization of the sign-LMS based adaptive filters,” \emph{IEEE Trans. Circuits Syst. I, Reg. Papers}, vol. 59, no. 9, pp. 2025--2033, Sep. 2012.
		
		\bibitem{VSS-LMS1}
		D. Bismor, K. Czyz, and Z. Ogonowski, “Review and comparison of variable step-size LMS algorithms,” \emph{Int. J. Acoust. Vib.}, vol. 21, no. 1, pp. 24-39, Mar 2016.
		
		\bibitem{VSS-LMS2}
		R. H. Kwong and E. W. Johnston, “A variable step size LMS algorithm,” \emph{IEEE Trans. Signal Process.}, vol. 40, no. 7, pp. 1633-1642, Jul 1992.
		
		\bibitem{VSS-LMS3}
		W. Ao, Y. Song, and C. Li, “A new variable step size LMS adaptive filtering algorithm,” in \emph{Proc. Int. Conf. Comput. Sci. Electron. Eng. (ICCSEE)}, Hangzhou, China, 2012, vol. 2, pp. 100-104.
		
		\bibitem{CG}
		D. Wu, “Scaled conjugate gradient digital cancellation algorithm study for 6G full duplex wireless communication system,” in \emph{Proc. IEEE Int. Wireless Commun. Mobile Comput. Conf. (IWCMC)}, 2024, pp. 1-6.
		
		\bibitem{RLS}
		Y. Jia, Z. Zhang, and L. Han, “Low complexity self-interference cancellation based on two-stage RLS filtering for full-duplex radios,” in \emph{Proc. IEEE Global Commun. Conf. (GLOBECOM)}, 2023, pp. 1-6.
		
		
		\bibitem{NN-base1} 
		M. Li, Y. Zhao, Q. Wang, et al., “RBF NN-enabled adaptive filter for any type of noise,” \emph{IEEE Trans. Neural Netw. Learn. Syst.}, early access, 2024. 
		
		\bibitem{NN-base2} 
		R. K. Agarwal, I. Hussain, and B. Singh, “Application of LMS-based NN structure for power quality enhancement in a distribution network under abnormal conditions,” \emph{IEEE Trans. Neural Netw. Learn. Syst.}, vol. 29, no. 5, pp. 1598–1607, May 2017.
		
		\bibitem{NN-base3} 
		F. Huang, S. Zhang, and W. X. Zheng, “Bayesian-learning-based diffusion least mean square algorithms over networks,” \emph{IEEE Trans. Neural Netw. Learn. Syst.}, vol. 35, no. 10, pp. 13217–13231, Oct 2024.
		
		\bibitem{theory-full-duplex}
		K. Komatsu, Y. Miyaji, H. Uehara and T. Matsumura, “Theoretical Investigation and Optimization of Power Amplifier Nonlinearity for In-Band Full-Duplex Radios,” in \emph{IEEE Transactions on Wireless Communications}, vol. 22, no. 5, pp. 3384--3396, May 2023.
		
		\bibitem{NONLMS1}
		X.~Yu, C.~Liu, and H.~Long, ``LMS-based digital nonlinear self-interference cancellation algorithm for SBFD systems,'' in \textit{Proc. IEEE Globecom Workshops (GC Wkshps)}, 2024, pp.~1--6.
		
		\bibitem{NONLMS2}
		H.~Lee and N.~Lee, ``Nonlinear and non-stationary self-interference cancellation for full-duplex wireless systems,'' in \textit{Proc. IEEE Int. Conf. Commun. (ICC)}, 2023, pp.~1370--1375.
		
		\bibitem{DeepOnet}
		L. Lu, P. Jin, G. Pang et al., “Learning nonlinear operators via DeepONet based on the universal approximation theorem of operators,” 
		\emph{Nat. Mach. Intell.}, vol. 3, pp. 218--229, 2021.
		
		\bibitem{Shift-DeepOnet}
		S. Lanthaler et al., “Nonlinear reconstruction for operator learning of PDEs with discontinuities,”
		\emph{arXiv:2210.01074} [math.NA], 2022. [Online]. 
		
		\bibitem{transformer}
		A. Vaswani et al., “Attention is all you need,” in \emph{Adv. Neural Inf. Process. Syst.}, vol. 30, 2017, pp. 5998-6008.
		
		\bibitem{transformer2}
		E. Calvello, S. Reich, and A. M. Stuart, “Continuum attention for neural operators,” \emph{arXiv:2406.06486 [cs.LG]}, 2024.
		
		\bibitem{PINO}
		Z. Li et al., “Physics-informed neural operator for learning partial differential equations,” 
		\emph{ACM/IMS J. Data Sci.}, vol. 1, no. 3, pp. 1--27, 2024.
		
		\bibitem{CNO}
		B. Raonic et al., “Convolutional neural operators for robust and accurate learning of PDEs,” in \emph{Adv. Neural Inf. Process. Syst.}, vol. 36, 2023, pp. 77187-77200.
		
		\bibitem{FNO}
		Z. Li et al., “Fourier neural operator for parametric partial differential equations,” in 
		\emph{Proc. Int. Conf. Learn. Represent. (ICLR)}, 2021. [Online]. Available: 
		\url{https://openreview.net/forum?id=c8P9NQVtmnO}
		
		\bibitem{FAN}
		Y. Dong et al., “Fan: Fourier analysis networks,” \emph{arXiv:2410.02675} [cs.LG], 2024. [Online].
		
		
		
		\bibitem{Balaji2026}
		R. Balaji, A. Kadiyala, D. Cabric, and S. Diggavi, ``Sample-Efficient Self-Interference Cancellation for In-Band Full Duplex Radios via In-Context Learning,'' in \emph{Proc. Workshop on AI and ML for Next-Generation Wireless Communications and Networking (AI4NextG), ICML}, 2026. [Online]. Available: \url{https://openreview.net/forum?id=QL5zUwojjK}

	\end{thebibliography}
\end{document}